\documentclass[conference,compsoc]{IEEEtran}
\IEEEoverridecommandlockouts

\usepackage{epsfig}
\usepackage{ifthen}
\usepackage{graphics}
\usepackage{epsfig}
\usepackage{setspace}
\usepackage[T1]{fontenc}
\usepackage{graphicx}
\usepackage{ifthen}
\usepackage{algorithm}
\usepackage[noend]{algorithmic}
\usepackage{makeidx}

\usepackage{ifpdf}

\usepackage{amsmath,latexsym,amssymb}
\usepackage{amsmath}
\usepackage{dsfont}
\usepackage{graphicx}
\usepackage{ifthen}
\usepackage{algorithm}
\usepackage[noend]{algorithmic}

\usepackage{ifpdf}

\newtheorem{theorem}{Theorem}
\newtheorem{lemma}[theorem]{Lemma}
\newtheorem{definition}[theorem]{Definition}

\def\rank{\mbox{\tt rank}}

\newcommand{\remove}[1]{}
\def\RANK{R}
\def\FAST{F}
\def\SLOW{S}
\def\LAYER{L}

\begin{document}

\title{{Latency Optimal Broadcasting \\in Noisy Wireless Mesh Networks}
\author{Qin Xin$^{\dag,\ddag}$, Yan Xia$^{\ast}$\\
$^\dag$Faculty of Science and Technology, University of the Faroe Islands, Faroe Islands \\Email: qinx@setur.fo ($^{\ddag}$Corresponding Author)\\
$^{\ast}$College of Computer Science and Electronics Engineering, Hunan University, Changsha, China\\Email: xiayan@hnu.edu.cn
}}

\maketitle

\begin{abstract}
Wireless mesh networking has been considered as an emerging communication paradigm
to enable resilient, cost-efficient and reliable services for the
future-generation wireless networks. We study here mainly on the minimum-latency
communication primitive of broadcasting (one-to-all communication) in
known topology WMNs, i.e., the size and the topology of the given Wireless Mesh Network (WMN) is known in advance.
A distinguished source mesh node in the WMN initially holds a "source" message and the
objective is to design a minimum-latency schedule such that the source message can be disseminated to all other mesh nodes. The
problem of computing a minimum-latency broadcasting schedule for a
given WMN is NP-hard, hence it is only possible to get a polynomial
approximation algorithm. In this paper, we adopt a new noisy wireless network
model introduced very recently by Censor-Hillel {\sl et al.} in [ACM PODC 2017, \cite{CHHZ17}].
More specifically, for a given noise parameter $p\in [0,1],$ any sender has a probability of $p$ of transmitting noise or any receiver of a single transmission in its neighborhood has a probability $p$ of receiving noise.

In this paper, we first propose a new asymptotically latency-optimal
approximation algorithm (under faultless model) that can complete single-message broadcasting task in $D+O(\log^2 n)$ time units/rounds in any WMN of
size $n,$ and diameter $D$. We then show this diameter-linear broadcasting algorithm remains robust under the noisy wireless network model and also improves the currently best known result in \cite{CHHZ17} by a $\Theta(\log\log n)$ factor.

In this paper, we also further extend our robust single-message broadcasting algorithm to $k$ multi-message broadcasting scenario and show it can broadcast $k$ messages in $O(D+k\log n+\log^2 n)$ time rounds. This new robust multi-message broadcasting scheme is not only asymptotically optimal but also
answers affirmatively the problem left open in \cite{CHHZ17} on the existence of an algorithm that is robust to sender and receiver faults and can broadcast $k$ messages in $O(D+k\log n + polylog(n))$ time rounds.\\
\end{abstract}


\section{Introduction}

Wireless Mesh Networking (WMN) is a highly promising network
architecture to converge the future-generation wireless networks. A
WMN has the dynamic self-organization, self-configuration and
self-healing characteristics; and additionally inherent flexibility,
scalability and reliability advantages. In a WMN, the mesh nodes can
communicate with each other via multi-hop routing or forwarding
\cite{Akyildiz}. There are two types of WMN with respect to the
mobility property, i.e. static mesh networks and mobile mesh
networks. The IEEE 802.11s mesh networks in Wireless Local Area
Networks (WirelessLAN) is a kind of WMN with static mesh nodes,
where the Access Points (APs) can communicate with each other via
multi-hop routing. Another example can be the WMN constructed by the
mesh routers with static topology. If the mesh nodes are equipped in
different moving objects, e.g. bicycles, buses and trains, the
network can be a kind of WMN with mobile mesh nodes. In this paper,
we focus on the WMN with static mesh nodes.

We consider the following model of a WMN:  an undirected connected
graph $G=(V,E),$ where $V$ represents the set of mesh nodes of the
WMN and $E$ contains unordered pairs of distinct mesh nodes, such
that  $(v,w)\in E$ iff the transmissions of mesh node $v$ can
directly reach mesh node $w$ and vice versa (the reachability of
transmissions is assumed to be a symmetric relation). In this case,
we say that the mesh nodes $v$ and $w$ are {\sl neighbors} in $G$.
Note that in a WMN, a message transmitted by a mesh node is always
potentially sent to all of its neighbors, which is the nature and
advantage of wireless communication.

The {\sl degree} of a mesh node  is the number of its neighbors. We use
$\Delta$ to denote the {\sl maximum degree} of the WMN, i.e.,
the maximum degree of any mesh node in the WMN. The {\sl size of the
network} is the number of mesh nodes $n = |V|$.

Communication in the WMN is synchronous and
consists of a sequence of communication steps/rounds.
In each step, a mesh node $v$ either transmits or listens. If $v$
transmits, then the transmitted message reaches each of its
neighbors by the end of this step.  However, a mesh node $w$
adjacent to $v$ successfully receives this message if only if in
this step $w$ is listening and $v$ is the only transmitting mesh
node among $w$'s neighbors under the classic faultless model. If mesh node $w$ is adjacent to a
transmitting mesh node but it is not listening, or it is adjacent to
more than one transmitting mesh nodes, then a {\sl collision} occurs
and $w$ does not retrieve any message in this step. Specifically, the classic
faultless model assumes that any message that is transmitted without collision
will be correctly received. Moreover,  we
assume that the collision is indistinguishable from the background
noise (that is, the mesh nodes do not have any collision detection
mechanism). Dealing with collisions is one of the main challenges in
efficient wireless communication.

In the noisy wireless network model introduced very recently by Censor-Hillel, Haeupler, Hershkowitz, and Zuzic in \cite{CHHZ17}, the classic graph-based faultless
model is augmented with random  faults.
More specifically, for a given noise parameter $p\in [0,1],$ any transmission may be
noisy with probability $p$ (called a sender fault), or a mesh node $u$ may receive a noise message with probability $p$ instead (called a receiver fault). Furthermore, the faults have been assumed to occur independently at each mesh node.

The two classical problems of information dissemination in the WMNs
are the {\sl single-message broadcasting} problem and the {\sl gossiping} problem.
The single-message broadcasting problem requires distributing a particular message
from a distinguished {\sl source} node to all other mesh nodes in
the WMN. In the gossiping problem, each mesh node $v$ in the network
initially holds a message $m_v$, and the aim is to distribute all
messages to all mesh nodes. A trade-off between the single-message broadcasting and the gossiping is multi-message broadcasting.  For all problems addressed above, the minimization of
the time needed to complete the task generally considers as the
efficiency criterion.

In the models considered here, the length of a communication schedule
is determined by the number of time rounds required to complete the
communication task. This means that we do not account for any
internal computation within individual mesh nodes.

Our schemes rely on the assumption that the communication algorithm
can use complete information about the WMN topology. Such an
assumption is acceptable since we investigate the communication
scenarios in static wireless mesh networks or classic radio networks here.
Such topology-based communication algorithms are useful whenever the underlying wireless network has
a fairly stable topology/infrastructure. As long as no changes occur
in the WMN topology during the execution of the algorithm, the tasks
of broadcasting and gossiping will be completed successfully.
%


\subsection{\bf Our results} In this paper, we first propose an
(efficiently computable) asymptotically latency-optimal approximation schedule
(under classic faultless model) that can complete single-message broadcasting task in $O(D+\log^2 n)$ time rounds/units in any WMN of
size $n,$ and diameter $D$. Note that computing a minimum-latency broadcasting
schedule is NP-hard, hence it is only possible to achieve
polynomial approximation algorithms. We then show that this diameter-linear broadcasting algorithm remains robust under the noisy wireless network model and it also improves the currently best known result in \cite{CHHZ17} by a $\Theta(\log\log n)$ factor.

We also further extend our robust single-message broadcasting algorithm to $k$ multi-message broadcasting scenario and show it can broadcast $k$ messages in $O(D+k\log n+\log^2 n)$ time rounds. This new robust multi-message broadcasting scheme is not only asymptotically optimal but also
answers affirmatively the problem left open in \cite{CHHZ17} on the existence of an algorithm that is robust to sender and receiver faults and can broadcast $k$ messages in $O(D+k\log n + polylog(n))$ time rounds.

\subsection{Related work} The work on
communication in known topology wireless networks was initiated in the
context of the single-message broadcasting problem. In \cite{CW87}, Chlamtac and
Weinstein prove that the broadcasting task can be completed in time
$O(D\log^2n)$ for every $n$-vertex wireless
network of diameter $D$. An $\Omega(\log^2n)$ time lower bound was proved for
the family of graphs of radius 2 by Alon {\sl et
al}  \cite{ABLP91}. In \cite{EK05}, Elkin and Kortsarz give an efficient deterministic
construction of a broadcasting schedule of length $D+O(\log^4n)$
together with  a $D+O(\log^3n)$ schedule for planar graphs.
Recently, G\k{a}sieniec, Peleg, and Xin \cite{GPX05} showed that a
$D+O(\log^3n)$ schedule exists for the broadcast task, that works in
{\em any} wireless network. In the same paper, the authors also provide
an optimal randomized broadcasting schedule of length $D+O(\log^2n)$
and a new broadcasting schedule using fewer than $3D$ time slots on
planar graphs. A $D+O(\log n)$-time broadcasting schedule for planar
graphs has been showed in \cite{MWX07} by Manne, Wang, and Xin.
Very recently, a $O(D+\log^2n)$ time deterministic
broadcasting schedule for any wireless network was proposed by Kowalski
and Pelc in \cite{KP-man}. This is asymptotically optimal unless
$NP \subseteq BPTIME(n^{{\cal{O}}({\log{\log n}})})$
\cite{KP-man}.
Nonetheless, for large $D$, in \cite{CMX06}, a $D+O(\frac{\log^3 n}{\log{\log n}})$ time
broadcasting scheme outperforms the one in \cite{KP-man},
because of the larger coefficient of the $D$ term hidden in the asymptotic notation
describing the time evaluation of this latter scheme.

Efficient single-message broadcasting algorithms for several special  types of
wireless network topologies can be found in Diks {\sl et al.}
\cite{DKP99}. In \cite{GPM03}, Gandhi, Parthasarathy and Mishra
claimed the NP-hardness of broadcasting in unit disk graphs and
constructed a broadcasting scheme with running time at most $648D$.
Very recently, the broadcasting time in unit disk graphs was further
reduced to $16D-15$ and $D+O(\log D)$ respectively by Huang et al.
\cite{HWJ+07}. For general wireless networks, however, it is known
that the computation of an optimal broadcast schedule is {NP-hard},
even if the underlying graph is embedded in the plane \cite{CK85,SH96}.  


Gossiping in wireless networks with known topology was first studied
in the context of the communication with messages of limited size,
by G\k{a}sieniec and Potapov in \cite{GP02}. They proposed several
optimal or close to optimal $O(n)$-time gossiping procedures for
various standard wireless network topologies, including lines,
rings, stars and free trees. In the same paper, an $O(n\log^2 n)$
gossiping scheme for general wireless network topology   is provided
and it is proved that there exists a wireless network topology in
which the gossiping (with unit size messages) requires $\Omega(n\log
n)$ time. In \cite{MX06}, Manne and Xin show the optimality of this
bound by providing an  $O(n\log n)$-time gossiping schedule with
unit size messages in any wireless radio network. The first work on
gossiping in known topology wireless networks with arbitrarily large
messages is \cite{GPX03}, where  several optimal gossiping schedules
are shown for a wide range of wireless network topologies. For
arbitrary topology of the wireless networks, an $O(D+\Delta\log n)$
schedule was given by G\k{a}sieniec, Peleg, and Xin in \cite{GPX05}.
Cicalese, Manne and Xin \cite{CMX06} provided a new
(efficiently computable) deterministic schedule that uses
$O(D+\frac{\Delta\log n}{\log{\Delta}-\log{\log n}})$ time units to
complete the gossiping task in any wireless network of maximum
degree $\Delta=\Omega(\log n)$. Later, Xin and Manne gave an asymptotically
optimal scheme with running time at $O(D+\frac{\Delta\log n}{\log{\Delta}})$ in \cite{XM09}.

The $k$ multi-message broadcasting problem had also been extensively studied.  Bar-Yehuda and Israeli \cite{Yehuda1989} proposed a $O((n + (k+D)\log n)\log \Delta)$-round algorithm, where $\Delta$ is the maximum node degree. A deterministic algorithm with running time in $O(n \log^4 n + k\log^3 n)$ rounds was shown by Chlebus {\sl et al.} \cite{chlebus2011efficient}. A nearly optimal $O(k\log n + D\log n/D + poly(\log n))$-round scheme was given by Ghaffari and Haeupler in \cite{ghaffari2013fast}.  Ghaffari {\sl et al.} proposed a scheme that can accomplish $k$ multi-message broadcasting task in $O(k\log n + D + \log^2 n)$ rounds if the topology is known in \cite{ghaffari2015randomized}.

Very recently, Censor-Hillel, Haeupler, Hershkowitz, and Zuzic in \cite{CHHZ17}
introduced a new wireless communication model that was called \emph{noisy radio network model}, in which the classic graph-based model is augmented with random faults. More precisely, for a constant fault parameter $p \in [0,1)$, every transmission may be noisy with probability $p$ (\emph{sender fault}), or a node $v$ that would otherwise receive a message with probability $p$ for noise (\emph{receiver fault}). Moreover, these faults occur independently at each node. It had been shown that while the Decay algorithm of Bar-Yehuda, Goldreich and Itai ~\cite{Yehuda1989} was robust to faults, the diameter-linear algorithm of G\k{a}sieniec, Peleg and Xin~\cite{GPX05} deteriorated considerably. A new randomized,  diameter-linear algorithm in the noisy radio network model had been proposed with running time $\Theta(D + \log n \log\log n(\log n + \log \frac{1}{\delta}))$ to complete the single-message broadcasting with a probability of at least $1 -\delta$. The work in \cite{CHHZ17} also described how to extend two robust single-message broadcasting schemes to the multi-message scenario, achieving throughputs of $\Omega\left(\frac{1}{\log n}\right)$ and $\Omega\left(\frac{1}{\log n \log\log n}\right)$ messages per round, respectively.

Latency-efficient communication schemes under the traditional physical
interference model had also been studied extensively in \cite{KLPP15,BP15,X09,X10,XW10,XWCF11}. Very recently, Xin and Xia proposed
a "noisy" physical interference model and showed some interesting latency-efficient gossiping schemes in \cite{XX17}.

\section{
Latency-optimal Broadcasting Schemes} \label{s:gossip}

In this section, we first present the idea of a new asymptotically
latency-optimal algorithm that generates a communication schedule for
completing the single-message broadcasting task in  $D+O(\log^2 n)$ time rounds under classic faultless model which is based on a non-trivial combination between a new scheme of the transmission pattern and the good properties of a {\em super gathering spanning tree} (SGST) in \cite{CMX06}. We then show this diameter-linear broadcasting algorithm remains robust under the noisy wireless network model and also improves the currently best known result in \cite{CHHZ17} by a $\Theta(\log\log n)$ factor. Finally, we also extend our robust single-message broadcasting algorithm to $k$ multi-message broadcasting scenario and show it can broadcast $k$ messages in $O(D+k\log n+\log^2 n)$. This new robust multi-message broadcasting scheme is not only optimal but also
answers affirmatively the problem left open in \cite{CHHZ17} on the existence of an algorithm that is robust to sender and receiver faults and can broadcast $k$ messages in $O(D+k\log n + polylog(n))$.

\subsection{Preliminaries}\label{rank}

For the convenience of our presentation as well as the self-containedness, we first recall the following recursive ranking procedure of nodes in a tree
(see \cite{CMX06}). Leaves have rank $1$. Next consider a mesh node
$v$ and the set $Q$ of its children and let $r_{max}$ be the maximum rank of the mesh nodes
in $Q$. Given a fixed integer parameter $2 \leq x \leq \Delta,$
if there are less than
$x$ mesh nodes in $Q$ of rank $r_{max}$ then set the rank of $v$ (e.g. rank$(v,x)$) to $r_{max}$,
otherwise set the rank of $v$ to $r_{max}+1$.

For an example, see Figure \ref{fig1}, where the same tree is ranked
with thresholds $x=2$ and $x=3$ respectively.\\



\begin{lemma}\label{s:rank}
Let  $T$ be a tree with $n$ nodes of maximum degree $\Delta.$
Then, $r_{max}^{[x]}
\leq \lceil\log_{x} n\rceil,$ for each $2 \leq x \leq \Delta,$
where $r_{max}^{[x]} = \max_{v \in T} \rank(v, x).$ (see \cite{CMX06}.)\\
\end{lemma}

For clarity of presentation, we reproduce some definitions from
\cite{CMX06}.

Given an arbitrary tree, we choose its central node $c$ as the root.
Then according to the hop distance from $c$, the mesh nodes in the
tree (rooted at $c$) are partitioned into consecutive layers
$\LAYER_i=\{v \mid dist(c,v)=i\},$ for $i=0,..,r$ where $0\le r\le
D$ is the radius of the tree. We denote the size of each layer
$\LAYER_i$ by $|\LAYER_i|.$

For a fixed value $x \geq 2,$ let
$\RANK_i(x)=\{v \mid \rank(v,x)=i\}$,
where $1\le i\le r_{max}^{[x]}.$

Based on the above rank sets, the mesh nodes can be divided into three
different types of transmission sets.\\

{
\begin{figure}[ht]
\epsfxsize=8cm
\center{\leavevmode \epsfbox{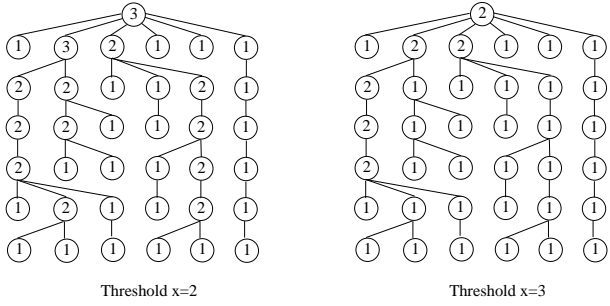}}
\caption{
\label{fig1}
\sf A tree of size $n=37$  ranked with $x=2$ (left) and  $x=3$ (right)}
\end{figure}
}

\begin{definition} \label{defi:partition}
The {\sl fast transmission set} is given by $\FAST^k_j=\{v
\mid v\in \LAYER_k\cap \RANK_j(2)$ and $parent(v)\in
\RANK_j(2)\}.$ Also define $\FAST_j=\bigcup_{k=1}^{D} \FAST^k_j$ and
$\FAST=\bigcup_{j=1}^{r_{max}^{[2]}
} \FAST_j$.\\
\end{definition}

\begin{definition}
The {\sl slow transmission set} is given by $\SLOW^k_j=\{v
\mid v\in \LAYER_k\cap\RANK_j(2)$ and $parent(v)\in \RANK_p(2),$ for
some $p>j;$ and $\rank(v,x)=\rank(parent(v),x), x>2\}$. Also define
$\SLOW_j=\bigcup_{k=1}^{D} \SLOW^k_j$ and
$\SLOW=\bigcup_{j=1}^{r_{max}^{[2]}
} \SLOW_j$.\\
\end{definition}


\begin{definition}
\noindent The {\sl super-slow transmission set} is given by
$SS^k_j=\{v \mid v\in \LAYER_k\cap \RANK_j(x)\ {\rm and}\
parent(v)\in \RANK_i(x), i>j\}.$ Accordingly, define
$SS_j=\bigcup_{k=1}^{D} SS^k_j$ and $SS=\bigcup_{j=1}^{r_{max}^{[x]}
}  SS_j$.\\
\end{definition}

Note that the above transmission sets define a partition of the node
set in that each mesh node $v$ only belongs to one of the transmission sets and
$V=F\bigcup S\bigcup SS.$\\
\subsubsection{The super gathering spanning tree}
In this subsection, we reproduce the definition of a {\em super
gathering spanning tree} (SGST) from \cite{CMX06}, which
plays an important role in our new broadcasting schemes in both the classic graph-based faultless and the noisy wireless network models described later.\\

\begin{figure}[htb]
\epsfxsize=8cm
\center{\leavevmode
\epsfbox{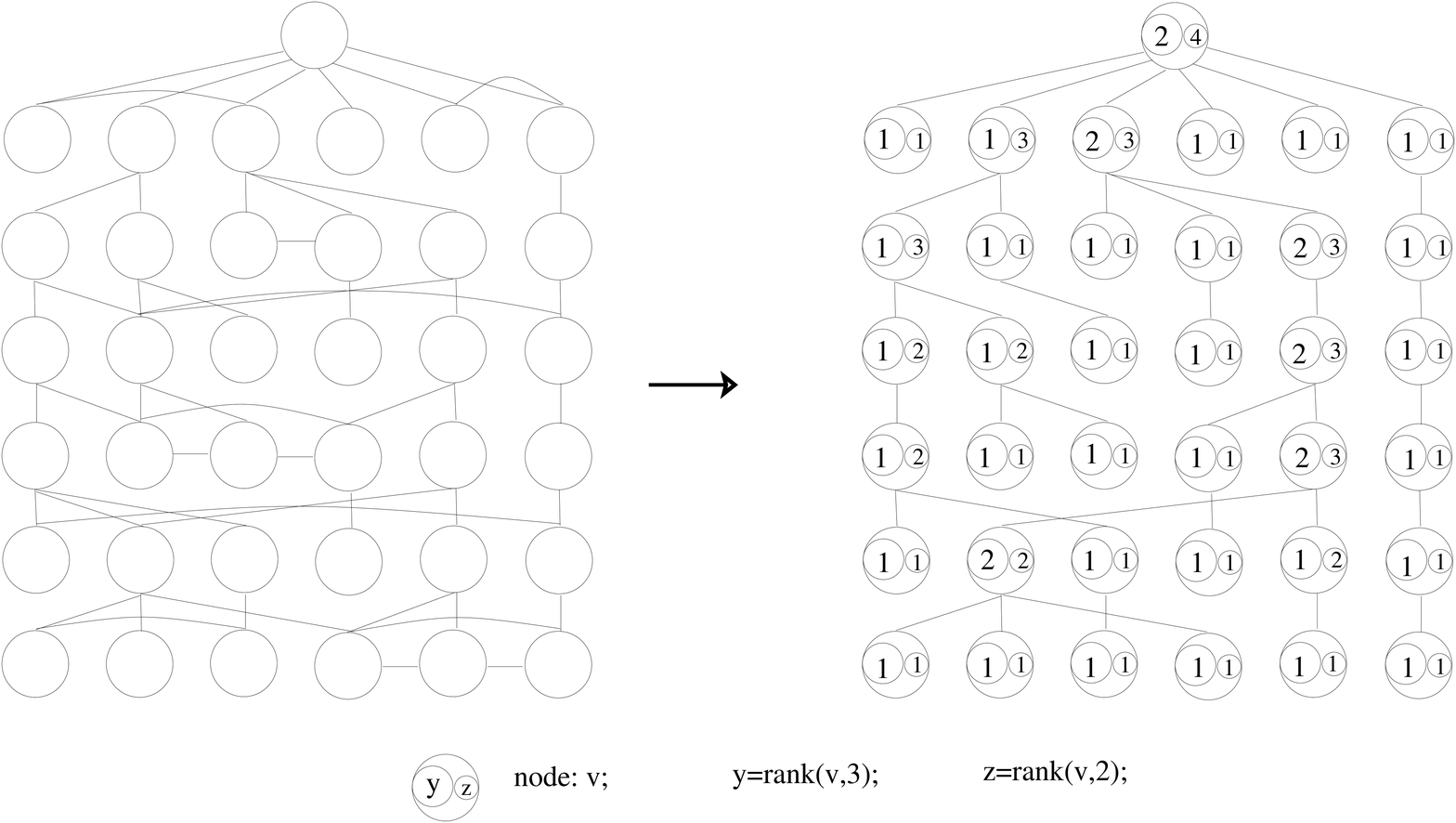}}
\caption{
\label{fig3}
\sf Construction of {a super gathering spanning tree}.}
\end{figure}

A {\em super gathering spanning tree} ({\em SGST}) for a
graph $G=(V,E)$ is any BFS spanning tree $T_G$\label{82} of $G,$
that satisfies:
\begin{description}
\item{(1)}
$T_G$ is rooted at the central node $c$ of $G$,
\item{(2)}
$T_G$ is ranked according to the ranking procedure from Section \ref{rank},
\item{(3)}
all mesh nodes in $\FAST^k_j$ of $T_G$ are able to transmit their
messages to their parents simultaneously without any collision, for
all  $1\le k\le D$ and $1\le j\le r_{max}^{[2]}\le\lceil\log
n\rceil$,
\item{(4)}
every mesh  node $v$ in $\SLOW^k_j\cap\RANK_i(x)$ of $T_G$ has the following
property: $parent(v)$ has at most $x-1$ neighbors in
$\SLOW^k_j\cap\RANK_i(x)$, for all
$i=1,2,...,r_{max}^{[x]}\le\lceil\log_x n\rceil$,
$j=1,2,...,r_{max}^{[2]}\le\lceil\log n\rceil$ and $k=1,...,D.$\\
\end{description}
An example has been shown in Figure \ref{fig3}.

We will use the following theorem for the analysis of our new
broadcasting scheme in Section \ref{sec_new_faultless}.\\

\begin{theorem}
For an arbitrary graph (e.g., an arbitrary WMN), there always exists an
$O(n^2\log n)$ time construction
of a $SGST$. (See \cite{CMX06}.)\\
\end{theorem}

\subsubsection{Decay Algorithm}\label{decay_a}
The classic Decay algorithm~\cite{Yehuda1989,CHHZ17} is to broadcast a single message from the source $s$ to all other nodes. The time rounds can be divided into phases of $O(\log n)$-round. During the $i^{th}$ round of each phase, where  $i \le O(\log n)$, each informed node broadcasts the message independently with probability $2^{-i}$.

The following Lemma had been shown in \cite{CHHZ17}.\\
\begin{lemma}\label{decay_lemma}
  If a node $v$ has an informed neighbor at the start of the phase, it becomes informed by the end of the phase with constant probability.\\
\end{lemma}
Consequently, we can also derive the following Lemma:\\
\begin{lemma}\label{lemma_bap}
In any bipartite graph, one partition (holding the message) can inform all nodes in another partition in one phase ($O(\log n)$ rounds), with constant probability.\\
\end{lemma}
The main Theorem of the time complexity of Decay algorithm (\cite{Yehuda1989}) states:
\begin{theorem}\label{decay_theorem}
In the faultless model, Decay algorithm broadcasts a single message in $O(D \log n + \log n(\log n + \log \frac{1}{\delta}))$ rounds with a probability of failure of at most $\delta$.\\
\end{theorem}
More precisely, it has been shown in \cite{CHHZ17}.
Fix a path $s = u_0, u_1, ..., u_l = v$ from the source $s$ to any node $v$ (the length $l$ of the path is at most the diameter $D$). At round $t$, let $\phi$ be the largest $i$ such that $u_i$ knows the message (initially, $\phi = 0$). After one phase ($O(\log n)$ rounds) $\phi$ either remains the same or increases by 1 with constant probability according to Lemma \ref{decay_lemma}. Therefore, after $O(D + \log n + \log \frac{1}{\delta})$ phases, the probability of failure can be bounded by a Chernoff bound:
\begin{align*}
  Pr[\Phi < l] < \exp\left(-\Omega\left(\log n + \log \frac{1}{\delta}\right)\right) .
\end{align*}
We can apply a union bound over all $n$ nodes and derive that the failure probability is at most $n \cdot \exp(-\Omega(\log n + \log \frac{1}{\delta})) < \exp(-\Omega(\log \frac{1}{\delta})) < \delta$. This is a very crucial property used in the analysis of our new broadcasting schemes.

\subsection{Optimal Single-Message Broadcasting Schedule in Faultless Model}\label{sec_new_faultless}

In this section, we show a new single-message broadcasting scheme of length
$D+O(\log^2 n)$ time rounds, which is asymptotically optimal. We adopt the randomized Decay algorithm in the Section \ref{decay_a} to replace the deterministic one used for the super-slow transmissions in \cite{CMX06}. Combining the new transmission pattern with the good properties of the {\em super gathering spanning tree}, we derive the claimed result.

In our single-message broadcast scheme, a super-gathering spanning tree rooted
at the {\sl source node} $s$ is used.
The broadcast message is now disseminated from the root
towards the leaves of the tree.

Similarly as the approach in \cite{CMX06}, we define $p(a)$ as the unique shortest path from the root $s$
to a leaf $a$. Note that the message does not necessarily follow the path
$p(a)$ and could actually even been delivered along non-shortest paths.
We can measure the delay from the time the message is already available
at some node $v$ on the path $p(a)$ to the time the message has already
reached the following node $w$ on the path (though not necessarily
via a transmission from $v$).

The path $p(a)$ can be though of as
consisting of several segments
$$p(a) ~=~ \langle p^F_1(a),p^S_1(a),p^{SS}_1(a),p^F_2(a),p^S_2(a),p^{SS}_2(a),
\ldots,$$
$$p^F_q(a),p^S_q(a),p^{SS}_q(a)\rangle~,$$
where each $p^F_i(a)$ is a segment consisting of fast transmission edges
(i.e., edges leading from $parent(v)$ to $v$ of $\rank(parent(v),2)=\rank(v, 2)$),
each $p^S_i(a)$ is an edge $(u,w)$ where $u$ is a node on layer $\LAYER_k$
for some $k$, $w$ is a node on layer $\LAYER_{k+1}$ and
$\rank(u,2)>\rank(w,2)$ and $\rank(u,x)=\rank(w,x)$.
We refer to such edges $(u,w)$ as slow transmission edges.
Further, each $p^{SS}_i(a)$ is an edge $(y,z)$ where $y$ is a node on layer $\LAYER_k$
for some $k$ and $z$ is a node on layer $\LAYER_{k+1}$ and $\rank(y,2)>\rank(z,2)$
and $\rank(y,x)>\rank(z,x)$.
We refer to such edges $(y,z)$ as super-slow transmission edges.
Note that some of the segments $p^F_i(a)$, $p^S_i(a)$ and $p^{SS}_i(a)$
may be empty.

The progress of the message dissemination  can be viewed as traversing the path $p(a)$
by alternating (flipping)
among chains $p^F_i(a)$ of fast transmission edges,
slow transmission steps over edges $p^S_i(a)$ and
super-slow transmission edges $p^{SS}_i(a).$

Next we describe the schedule governing these transmissions.
Consider a node $v$ with $1\le \rank(v,2)\le r_{max}^{[2]}$ on BFS
layer $\LAYER_i$ with a child $w$ of the same rank at the next BFS
layer. Then $v$ can perform a fast transmission to $w$ in a time
step $t$ satisfying $t\equiv i+9j \bmod 9r_{max}^{[2]}$, where
$j=\rank(v,2)$.
The slow transmissions at the BFS layer $\LAYER_i$ are performed
in the time steps $t$ satisfying $t\equiv i+3\bmod 9$.
The super-slow transmissions at the BFS layer $\LAYER_i$ are performed
in the time steps $t$ satisfying $t\equiv i+6\bmod 9$.
This way, the fast, the slow and the super-slow transmissions
at any BFS layer are separated by three units
of time. Thus, there are no collisions between the fast, the slow and super-slow
transmissions at the same BFS layer.
Moreover, there cannot be conflicts
between transmissions coming from different BFS layers either.
In fact, at any time step, transmissions are performed on BFS layers at
distances that are multiples of 3 apart.

When the message arrives at the first node $v$
of a fast segment $p^F_i(a)$ of the route (with a particular rank),
it might wait for as many as $9r_{max}^{[2]}=O(\log n)$ time
steps before being
transmitted to the next BFS layer. However,
it will then be forwarded through the fast segment $p^F_i(a)$
without further delays.

Once reaching the end node $u$ of the fast segment $p^F_i(a)$,
the message has to be transmitted from some node on $u$'s BFS layer
to the next node $w$ on $p(a)$, which has $\rank(u,2)>\rank(w,2)$
and $\rank(u,x)=\rank(w,x)$,
using a slow transmissions mechanism.
For slow transmissions, the algorithm uses the $x$ transmissions to progress
distance one on $p(a)$ due to the property of the $SGST$. Note that the transmission patterns for the fast and the slow transmissions are identical as the one in \cite{CMX06}.

Once reaching the end node $y$ of the slow segment $p^S_i(a)$,
the message has to be transmitted from some node on $y$'s BFS layer
to the next node $z$ on $p(a)$, which has $\rank(y,2)>\rank(z,2)$
and $\rank(y,x)>\rank(z,x)$,
using a super-slow transmissions mechanism.
For super-slow transmissions, our algorithm uses
the Decay algorithm mentioned in Section \ref{decay_a} with a $O(\log n)$ transmission rounds in each phases.
By Lemma \ref{lemma_bap}, the Decay algorithm allows to move uniform information from one partition
of a bipartite graph of size $n$
(here, an entire BFS layer $\LAYER_j$ of the tree)
to the other (here, the next layer $\LAYER_{j+1}$)
in time $O(\log n)$ with constant probability. Since the path can be decomposed into at most $O(\log_x n)$ super-slow edges. By Theorem \ref{decay_theorem}, the message successfully traverses all of the super-slow edges after
$O(\log n(\log_x n + \log \frac{1}{\delta}))$ transmission rounds with a probability of at least $1-\frac{\delta}{n}$.

By virtue of the above observations we can bound
the total time required for the
broadcast the single source message to reach a leaf $a$ as follows.
Let $D_i$, for $1\le i\le r_{max}^{[2]}$, denote the length of $p^F(a)$,
the $i$th fast segment of the route $p(a)$ used by the broadcast message
that has reached $a$.
Thus the time required to communicate $a$ is bounded by
$O(\log n)+D_{1}+\ldots+O(\log n)+D_{r_{max}^{[2]}} \le D+O(\log^2 n)$ (with probability $1$)
for the fast transmissions plus $r_{max}^{[2]}\cdot O(x)=O(x\log n)$ (with probability $1$) for the
slow transmissions and
$O(\log n(\log_x n + \log \frac{1}{\delta}))$ with a probability of at least $1-\frac{\delta}{n}$
for the super-slow transmissions, yielding a total
of $D+O(\log^2 n+x\cdot\log n+\frac{\log^2 n}{\log x})$ rounds with a probability
of at least $1-\frac{\delta}{n}$. Combing with the union bound over all nodes, we can summarize our findings in the following theorem.\\

\begin{theorem}
In the faultless model, we can spreads a single message in
$D+O(\log^2 n+x\cdot\log n+\log n(\log_x n + \log \frac{1}{\delta}))$ rounds with a probability
of failure of at most $\delta$. In particular,
by setting $x=\Theta(\log n),$ we obtain the
bound $D+O(\log^2 n)$.
\end{theorem}

\subsection{Optimal Single-Message Broadcasting in Noisy Model}

In this section, we show how we adapt the single-message broadcasting algorithm
we derived at Section \ref{sec_new_faultless} from the faultless setting to the sender or receiver faults setting (the noisy network model) in order to obtain robust single-message broadcast scheme. Our new robust scheme is based on the framework in \cite{CHHZ17},

As in the broadcasting scheme at Section \ref{sec_new_faultless}, a SGST is constructed from the source node $s$. The communication process is  split
into consecutive blocks of 9 time rounds each.
The first 3 rounds of each block are used for fast transmissions
from the set $\FAST$, the middle 3 rounds are reserved for slow
transmissions from the set $\SLOW$ and  the remaining 3 are used
for super-slow transmissions of the mesh nodes
from the set $SS.$
We use 3 rounds of time for each type of transmission
in order to prevent collisions between
neighboring BFS layers.

During super-slow transmission rounds, a standard Decay algorithm (see Section \ref{decay_a}) is performed on all nodes. These rounds are meant to push the message from one fast stretch or one slow transmission to the next.

During slow transmission rounds, we replace the deterministic approach of faultless broadcasting by the standard Decay algorithm. Note also the number of informed nodes who are competing the transmission at any node (associating on a slow transmission edge) can be bounded by the ranking parameter $x$ due to the properties of the SGST.

During fast transmission rounds, we adopt the framework from \cite{CHHZ17} but modify the transmission pattern for the fast stretches. First, partition the nodes of each fast stretch into blocks of size $S := \Theta(\log \log n)$ (all the blocks have size $\Theta(\log\log n)$, except possibly the last one). The procedure of {broadcasting on a block} has been defined in the following way: a node $v$ with $1\le \rank(v,2)\le r_{max}^{[2]}$ on BFS
layer $\LAYER_i$ (with a child $w$ of the same rank at the next BFS
layer) can perform a fast transmission to $w$ in a time
step $t$ satisfying $t\equiv i+9j \bmod 9r_{max}^{[2]}$, where
$j=\rank(v,2)$. The procedure continues on for $c \cdot S = \Theta(\log \log n)$ rounds for some sufficiently large constant $c$. Note that it has been stated in \cite{CHHZ17} the probability that a message that is in a broadcasting block in the beginning fails to exit the block is at most $\frac{1}{\log^{c'} n}$ for a constant $c'$ which  can be set as large as needed (by increasing the round multiplier $c$).

We use the same concept "supernode" from \cite{CHHZ17} to contract the nodes in a block into one. A broadcast on this supernode corresponds to the block-broadcast procedure described in the last paragraph and {"superrounds"} on this graph correspond to $\Theta(\log \log n)$ rounds in the original graph.

Similarly, we can define ranks, BFS levels and fast transmission supernodes in the same way as in the original graph we defined in the last Section \ref{sec_new_faultless}. The algorithm on the contracted graph is as follows: at round $t$, a fast supernode with level $i$ and rank $j$ broadcasts if $t\equiv i+9j \bmod 9r_{max}^{[2]}$; the slow transmissions at the BFS layer $\LAYER_i$ are performed in the time rounds $t$ satisfying $t\equiv i+3\bmod 9$; and the  super-slow transmissions at the BFS layer $\LAYER_i$ are performed
in the rounds $t$ satisfying $t\equiv i+6\bmod 9$.
Consequently, the fast, the slow and the super-slow transmissions
at any BFS layer are separated by three rounds
of time. Thus, there are no collisions between the fast, the slow and super-slow
transmissions at the same BFS layer.
Moreover, there cannot be conflicts between transmissions coming from different BFS layers either. In fact, at any time step, transmissions are performed on BFS layers at distances that are multiples of 3 apart.

Consider any SGST-path $p(a)$ on the contracted graph
$$p(a) ~=~ \langle p^F_1(a),p^S_1(a),p^{SS}_1(a),p^F_2(a),p^S_2(a),p^{SS}_2(a),
\ldots,$$
$$p^F_q(a),p^S_q(a),p^{SS}_q(a)\rangle~,$$
from $s$ to another node $a$ and note that it has at most $r_{max}^{[2]} = O(\log n)$ fast stretches and at most $r_{max}^{[2]} = O(\log n)$ slow transmission edges, and at most $r_{max}^{[x]} = O(\log_x n)$ super-slow transmission edges.

Assuming a message is on a super-slow transmission edge, by following the standard Decay algorithm,  during the next $\Theta(\log n)$ rounds it is transmitted along that edge with constant probability according to Lemma \ref{decay_lemma}. Given the fact that there are only $O(\log_x n)$ such edges, a Chernoff bound gives us that after $O(\log n(\log_x n + \log \frac{3}{\delta}))$ such time rounds, the message is transmitted along all the super-slow edges on $p(a)$ with probability at least $1 - \frac{\delta}{3n}$. Applying a union bound over all $n$ nodes gives that the failure probability is at most $n \cdot \exp(-\Omega(\log n + \log \frac{3}{\delta})) < \exp(-\Omega(\log \frac{3}{\delta})) < \frac{\delta}{3}$.

Similarly, we assume a message is on a slow transmission edge. By the construction and the properties of a SGST, we know that the number of informed neighboring nodes can be bounded by $x-1$. Therefore, the graph induced by the slow transmission edges together with the corresponding incident nodes can be colored by $x$ ($\Theta(x)$). We can now consider all nodes with same colour as one single node since they perform in an identical transmission manner.
By following the standard Decay algorithm,  during the next $\Theta(\log x)$ rounds it is transmitted along that edge with constant probability according to Lemma \ref{decay_lemma}. Given the fact that there are only $O(\log n)$ such edges, we can use a Chernoff bound to show that after $O(\log x(\log n + \log \frac{3}{\delta}))$ such time rounds, the message is transmitted along all the slow edges on $p(a)$ with probability at least $1 - \frac{\delta}{3x}$. Applying a union bound over all $x$ nodes ($\Theta(x)$ coloring scheme) gives that the failure probability is at most $x \cdot \exp(-\Omega(\log x + \log \frac{3}{\delta})) < \exp(-\Omega(\log \frac{3}{\delta})) < \frac{\delta}{3}$.

Finally, we start to counting the number of time rounds that a message spends on fast stretches (during fast transmission rounds). Note that from the design of the algorithm (transmission patterns) and the properties constructed by the SGST no two broadcasting nodes ever interfere with each other. Therefore, the only failures occurs from constant probability faults. We follow the similar strategies and analysis from \cite{CHHZ17} but more smartly use the properties of the SGST. Combing with a better probability setting (e.g., by an appropriate constant $c$), we improves a $\Theta(\log\log n)$ factor from the best robust single-message broadcasting scheme proposed in \cite{CHHZ17}.

For the sakes of clarification and comparison, we use the same concepts from \cite{CHHZ17}. Call a fast transmission node from the path $p(a)$ a barrier if its BFS level is divisible by $S$ and call a message active if it is on a fast transmission stretch and the node it is currently at is broadcasting. We also define a new concept "connector". Call a slow transmission edge from the path
$p(a)$ a connector if it connects two fast transmission stretches. A connector will be used to glue two separate fast stretches together (with extra small overhead, e.g., $O(\log\log n)$ rounds by setting $x=\log n$) to minimize the total number of the fast stretches. Note that a message that enters a fast stretch has to wait $r_{\max}^{[2]}cS = O(\log n \log\log n)$ rounds until it becomes active. Once it is active, we now analyze its behavior during the next $cS = O(\log\log n)$ rounds. The message can either exit the fast stretch, remain active (reaching the next barrier) or become inactive (failing $(c-1)S$ out of $cS$ transmissions). We now follow the framework in \cite{CHHZ17} and bound the probability of becoming inactive by using a different probability setting, which is at most $\frac{1}{\log^3 n\log\log n}$ by Chernoff with an appropriate constant $c$. Every time a message becomes inactive, it waits $O(\log n \log\log n)$ rounds before it becomes active.

Let $d_1, d_2, ..., d_q$ be the lengths of the fast stretches in the path $p(a)$. Combining the properties of the SGST with the power and functionality of connectors, we can bound $q \le r_{\max}^{[x]}=O(\log_x n)$. By choosing $x=\log n$, $q \le \Theta(\frac{\log n}{\log\log n})$. When a message is active, it traverses the paths in at most $\sum_{i=1}^q \lceil d_i/S \rceil \cdot cS = \sum_{i=1}^q O(d_i + 1) = O(D + \frac{\log n}{\log\log n})$ rounds. Note that the extra small overheads at the connectors can be bounded in the slow transmission rounds by $O(\log n\log\log n)$ in total with a failure probability at most  $\frac{\delta}{3}$.
  The number of rounds it takes for a message to become active is at most\\
  \begin{align*}
    &\left (q + \frac{T}{cS} Pr[\text{msg inactive in $cS$ rounds}] \right )\cdot O(\log n \log\log n)\\
    &\le O(\log^2 n) + \frac{T}{\Theta(\log \log n)} \frac{O(\log n\log\log n)}{\log^3 n\log\log n}&\\
    &= O(\log^2 n) + \Theta \left (\frac{T}{\log^2 n\log\log n} \right )\\
  \end{align*}
  where $T$ is the total length of the robust single-message broadcasting scheme. The $q$ term comes from becoming active each time a message enters a fast transmission stretch under consideration of the help from the connectors. The $\frac{T}{cS}$ accounts for the possibility of a message becoming inactive in between barriers. Note that the work in \cite{CHHZ17} requires
  $O(\log^2 n\log\log n)+\Theta\left(\frac{T}{\log^2 n}\right)$ number of time rounds to handle the fast stretches. Consequently, our new scheme saves a
  $\Theta({\log\log n})$ factor to complete the message broadcasting in the fast stretches. Under a same Chernoff bound as \cite{CHHZ17} together with the approaches we used to handle slow-transmission edges and super-slow transmission edges, we can prove that if $T = \Theta(D + \log n (\log n + \log \frac{3}{\delta}))$, the message gets passed along the path with a probability of at least $1 - \frac{\delta }{3n}$. Similarly, the union bound gives that the failure probability is at most $\frac{\delta}{3}$.

  Putting together the behavior during the fast, slow, super-slow transmission  rounds gives that the protocol forwards the message from the source to all other nodes in the claimed number of rounds with probability at least $1 - \delta$.

Consequently, we specify our main results in this section in the following Theorem.\\

\begin{theorem}
Our robust broadcasting scheme spreads a single message in $O(D + \log n(\log n + \log \frac{1}{\delta}))$ rounds with a probability of failure of at most $\delta$ if sender or receiver faults occur with probability $p$.
\end{theorem}

\subsection{Robust Algorithms for Multi-Message Broadcast}
Haeupler in \cite{H2011} states that the interesting feature of single-message broadcasting algorithms that are robust to sender failures is that they can be used in a black-box manner to transmit $k$ messages with random linear network coding. Based on the same conditions in \cite{CHHZ17} to satisfy the requirements to use random linear network coding,  we state the results that can be achieved and refer the reader to \cite{CHHZ17} for details.\\

\begin{theorem}
Our robust single-message broadcasting scheme with random linear network coding can broadcast $k$ messages in $O(D + k\log n + \log^2 n)$ rounds if sender or receiver faults occur with constant probability. It follows that any topology has a coding throughput of $\Omega\left(\frac{1}{\log n}\right)$.
\end{theorem}

\section{Conclusion}
In this paper, we propose asymptotically latency-optimal schedules under both classic faultless and noisy wireless network models that can complete single-message broadcasting task in $O(D+\log^2 n)$ time units in any WMN of
size $n,$ and diameter $D$, which also improves the currently best known result in \cite{CHHZ17} by a $\Theta(\log\log n)$ factor. We also show how to extend our robust single-message broadcasting algorithm to $k$ multi-message broadcasting scenario and achieve a time bound on $O(D+k\log n+\log^2 n)$. This new robust multi-message broadcasting scheme is not only asymptotically optimal but also
answers affirmatively the main problem left open in \cite{CHHZ17}. We hope our work can stimulate the further research on reliable and robust communication in wireless networks.

 
{

}


\begin{thebibliography}{6}


{
\bibitem{Akyildiz}
I.F. Akyildiz, X. Wang and W. Wang.
Wireless mesh networks: a survey.
{\em Elsevier Computer Networks}, vol. 47, no. 4, pp.
445-487, 2005.}


\bibitem{ABLP91}
N.~Alon, A.~Bar-Noy, N.~Linial and D.~Peleg.
A lower bound for radio broadcast.
{\em J. Computer and System Sciences} 43, (1991), 290 - 298.


\bibitem{BP15}
Barenboim L., Peleg D. \emph{Nearly Optimal Local Broadcasting in the SINR Model with Feedback}. In: Scheideler C. (eds) Structural Information and Communication Complexity. SIROCCO 2015. Lecture Notes in Computer Science, vol 9439.

\bibitem{haeuplersodap1843}
Noga Alon, Mohsen Ghaffari, Bernhard Haeupler, and Majid Khabbazian.
\newblock Broadcast throughput in radio networks: Routing vs. network coding.
\newblock \emph{ACM-SIAM Symposium on Discrete Algorithms (SODA) 2014}, pages
  1831--1843, 2014.


\bibitem{Yehuda1989}
R.~Bar-Yehuda and A.~Israeli.
\newblock Multiple communication in multi-hop radio networks.
\newblock In \emph{the proceedings of the 8th Annual ACM Symposium on Principles
  of Distributed Computing, PODC 1989}, pages 329--338, New York, NY, USA, 1989.


\bibitem{CHHZ17}
B. Haeupler, E. Hershkowitz, G. Zuzic and K. Censor-Hillel.
Broadcasting in Noisy Radio Networks. {\em in the proceedings of the 36th ACM Symposium on Principles of Distributed Computing, PODC 2017}, July 25-27, 2017, Washington, DC, USA. (https://arxiv.org/abs/1705.07369)

\bibitem{CK85}
I.~Chlamtac and S.~Kutten.
On broadcasting in radio networks-problem analysis and protocol design.
{\em IEEE Trans. on Communications} 33, (1985), pp. 1240-1246.

\bibitem{CMX06}
F.~Cicalese, F.~Manne, and Q.~Xin: Faster Centralized Communication
in Radio Networks. {\em Algorithmica}, 54(2):226-242, 2009.

\bibitem{CW87}
I.~Chlamtac and O.~Weinstein.
The wave expansion approach to broadcasting in multihop radio networks.
{\em IEEE Trans. on Communications} 39, (1991), pp. 426-433.

\bibitem{chlebus2011efficient}
Bogdan~S Chlebus, Dariusz~R Kowalski, Andrzej Pelc, and Mariusz~A Rokicki.
\newblock Efficient distributed communication in ad-hoc radio networks.
\newblock In \emph{International Colloquium on Automata, Languages, and
  Programming}, pages 613--624. Springer, 2011.

\bibitem{DKP99}
K.~Diks, E.~Kranakis, D. Krizanc and A.~Pelc. The impact of
information on broadcasting time in linear radio networks. {\em
Theoretical Computer Science}, 287, (2002), pp. 449-471.


\bibitem{EK05}
M.~Elkin and G.~Kortsarz.
Improved broadcast schedule for radio networks.
{\em Proc. 16th ACM-SIAM Symp. on Discrete Algorithms}, 2005, pp. 222-231.



\bibitem{ghaffari2013fast}
Mohsen Ghaffari and Bernhard Haeupler.
\newblock Fast structuring of radio networks large for multi-message
  communications.
\newblock In \emph{International Symposium on Distributed Computing}, pages
  492--506. Springer, 2013.

\bibitem{ghaffari2015randomized}
Mohsen Ghaffari, Bernhard Haeupler, and Majid Khabbazian.
\newblock Randomized broadcast in radio networks with collision detection.
\newblock \emph{Distributed Computing}, 28\penalty0 (6):\penalty0 407--422,
  2015.


\bibitem{GP02}
L.~G\k{a}sieniec and I.~Potapov, Gossiping with unit messages in
known radio networks. {\em Proceedings of 2nd IFIP TCS}, 2002, pp.
193-205.

\bibitem{GPM03}
R. Gandhi, S. Parthasarathy, A. Mishra. Minimizing broadcast latency
and redundancy in ad hoc networks. {\sl Proc. of the 4th ACM
International Symposium on Mobile Ad Hoc Networking and Computing},
MobiHoc 2003, pp. 222-232.

\bibitem{GPX03}
L.~G\k{a}sieniec, I.~Potapov and Q.~Xin. Efficient gossiping in
known radio networks. {\sl Proc. 11th SIROCCO}, 2004, LNCS 3104, pp.
173-184.


\bibitem{GPX05}
L.~G\k{a}sieniec, D.~Peleg and Q.~Xin. Faster communication in known
topology radio networks. {\sl Proc. 24th Annual ACM SIGACT-SIGOPS
PODC}, 2005, pp. 129-137. Also in Distributed Computing 19(4): 289-300 (2007).



\bibitem{H2011}
Bernhard Haeupler. Analyzing network  coding gossip made easy. In {\sl Proceedings of the 43rd Annual ACM symposium on Theory of Computing,}, STOC 2011, pp. 293-302.

\bibitem{HWJ+07}
S.C.H. Huang, P.J. Wan, X. Jia, H. Du, and W. Shang. Minimum-Latency
broadcast scheduling in wireless ad hoc networks. {\sl Proc. of the
26th Annual IEEE Conference on Computer Communicaitions}, IEEE
INFOCOM 2007, pp. 733-739.
 

\bibitem{KLPP15}
E. Kantor, Z. Lotker, M. Parter and D. Peleg, {\sl The Minimum Principle of SINR: A Useful Discretization Tool for Wireless Communication,} 2015 IEEE 56th Annual Symposium on Foundations of Computer Science, Berkeley, CA, 2015, pp. 330-349.

\bibitem{KP-man}
D.~Kowalski and A.~Pelc. Optimal deterministic broadcasting in known
topology radio networks. {\em Distributed  Computing}, 19(3):
185-195 (2007).


\bibitem{MWX07}
F.~Manne, and  Q.~Xin. Faster radio broadcast in planar graphs.
{\em Journal of Networks,} 3(2): 9-16, 2008.

\bibitem{MX06}
F.~Manne and Q.~Xin.
Optimal gossiping with unit size messages in known radio networks.
{\em Proc. 3rd Workshop on Combinatorial and Algorithmic Aspects of Networking}, LNCS 4235, pp. 125-134.




\bibitem{SH96}
A.~Sen and M.L.~Huson. A new model for scheduling packet radio
networks. {\em Proc. 15th Joint Conf. of IEEE Computer and
Communication Societies}, INFOCOM 1996, pp. 1116-1124.

\bibitem{X09}
Q. Xin, J. Xiang. Joint admission control, channel assignment and power
allocation in cognitive radio cellular networks, in: {\sl proc. of
the IEEE International Conference on Mobile Ad-hoc and Sensor
Systems,} IEEE MASS 2009, pp. 294-303.

\bibitem{X10}
Q. Xin.  Minimum-Latency Communication in Wireless Mesh Networks Under Physical Interference Model. {\em Proc. IEEE 45th International Conference on Communications,} ICC 2010, pp. 1-6.

\bibitem{XM09}
Q. Xin, F. Manne, and X. Yao. Latency-optimal communication in wireless mesh
networks. In {\em Theoretical Computer Science,} 528: 79-84 (2014).

\bibitem{XW10}
Q. Xin and Y. Wang.  Latency-efficient M2M Multicasting in Wireless
Mesh Networks Under Physical Interference Model. {\em Proc. IEEE
Wireless Communications and Networking Conference,} WCNC 2010.

\bibitem{XWCF11}
Q. Xin, X. Wang, J. Cao, W. Feng.
Joint Admission Control, Channel Assignment and QoS Routing for Coverage Optimization in Multi-Hop Cognitive Radio Cellular Networks. in: {\sl proc. of
the IEEE International Conference on Mobile Ad-hoc and Sensor
Systems,} IEEE MASS 2011, pp. 55-62.

\bibitem{XX17}
Q. Xin and Y. Xia.
Minimum-Latency Communication in Wireless Mesh Networks
Under Noisy Physical Interference Model. in: {\sl proceedings of
the 9th IEEE International Conference on Wireless Communications and Signal Processing,} IEEE WCSP 2017, to appear.

\end{thebibliography}
\end{document}